\documentclass[a4paper]{jpconf}
\usepackage{graphicx}

\usepackage{epsfig}
\usepackage{graphicx}
\usepackage{dcolumn}
\usepackage{amsmath}
\usepackage{amssymb}
\usepackage{enumerate}
\usepackage{subfigure}    
\usepackage{multirow} 
\usepackage{cite}

\begin{document}
\title{Long-baseline super-beam experiments in Europe within LAGUNA}

\author{Pilar Coloma$^{{\rm a}\dagger}$ , Tracey Li$^{\rm b}$, Silvia Pascoli$^{\rm c}$}
\address{$^{\rm a}$Departamento de F\'isica Te\'orica \& Instituto de F\'isica Te\'orica, Universidad Aut\'onoma de Madrid. 28049 Cantoblanco, Madrid (Spain)}
\address{$^{\rm b}$Instituto de F\'isica Corpuscular, Edificio Institutos de Investigaci\'on, Paterna, Valencia E-46071  (Spain)}
\address{$^{\rm c}$IPPP, Department of Physics, Durham University, Durham DH1 3LE (United Kingdom)}
\address{$^{\dagger}$ Talk given at NuFact'11 (Geneva, August 1-6, 2011).}
\ead{$^{\rm a}$p.coloma@uam.es, $^{\rm b}$tracey.li@ific.uv.es, $^{\rm c}$silvia.pascoli@durham.ac.uk}



\begin{abstract}
We explore the physics reach of several possible configurations for a Super-Beam experiment in Europe, focusing on the possibilities of discovering $\theta_{13}$, CP violation in the leptonic sector and the ordering of neutrino mass eigenstates.
We consider the three different detector technologies: Water \v{C}erenkov, Liquid Argon and Liquid Scintillator, and seven possible sites in Europe which would be able to host such a detector underground. The distances to these sites from CERN, where the beam would be originated, go from 130 km to 2300 km. The neutrino flux is optimized in each case as to match the first oscillation peak for each of the baselines under consideration. 
We also study the impact of several experimental factors in the performance of each detector technology. These include the influence of the spectral information, the rejection efficiencies for the neutral-current backgrounds, the ratio between running times in neutrino and antineutrino modes and the systematic uncertainties on the signal and backgrounds, among others. 
Contribution to NUFACT 11, XIIIth International Workshop on Neutrino Factories, Super beams and Beta beams, 1-6 August 2011, CERN and University of Geneva (Submitted to IOP conference series).

\end{abstract}


\section{Introduction and simulation details}

Future long-baseline neutrino oscillation experiments are being designed to measure the neutrino oscillation parameters $\theta_{13},\,\delta$ and the mass hierarchy. We evaluate the performance of such an experiment within LAGUNA~\cite{Angus:2010sz}, optimizing and comparing the performances of seven different baselines and three detector technologies. In particular, we focus on a Super-Beam (SB) experiment based at CERN. SB fluxes have been optimized for each baseline according to the first oscillation peak~\cite{Longhin:2010zz}. In the lower energy configuration for the beam (for $L=130$ km) the fluxes have been obtained assuming $0.56\times 10^{23}$ Protons on Target (PoT) per year, with an energy of $4.5$ GeV. We assume 2 years of $\nu$ running and 8 years of $\bar{\nu}$ running in this case, as in Ref.~\cite{Campagne:2006yx}. Higher energy fluxes (for setups with $L>130$ km) correspond to the CERN high-power PS2 configuration~\cite{Rubbia:2010fm}: $3\times10^{21}$ PoT per year, with an energy of $50$ GeV. We assume $5+5$ years in this case, though, since we have checked that an asymmetric configuration is only marginally beneficial only for the $\theta_{13}$ discovery potential of the experiment.

We use the GLoBES software~\cite{globes1,globes2} in all our simulations. Marginalization has been performed over solar and atmospheric parameters, according to their present best-fits and $1\sigma$ errors~\cite{Schwetz:2011qt}. Whenever we marginalize over $\theta_{13}$ and $\delta$ we leave them completely free. A conservative $5\%$ uncertainty has been taken over the PREM profile~\cite{matter1}. Full details used for the simulation of the detectors response can be found in \cite{ourpaper}. Information about the Liquid Argon (LAr) detector has been obtained from Refs.~\cite{LBNEreport,Barger:2007jq} with migration matrices provided by L.~Esposito and A.~Rubbia. Information about the Liquid Scintillator (LS) detector has been obtained from Refs.~\cite{randolph,randolph2,Wurm:2011zn}. The rejection capability for the Neutral-Current (NC) events at the LS detector is currently under study. Therefore, we have considered that a certain percentage of the total amount of unoscillated NC events will be misidentified as Charged-Current (CC) events and we have varied it within $0.5\%$ and $5\%$. Finally, for the WC detector we have considered two different configurations: for the lower energy beam configuration we have used the same details as in Ref.~\cite{Campagne:2006yx}; however, for the higher energy setups we have followed Refs.~\cite{LBNEreport,brajesh}. The migration matrices in this case have been kindly provided by the LBNE collaboration~\cite{Lisa}, as well as the rejection efficiencies for the NC background events. For all detectors we assume the level of the intrinsic electron background to be the same as the signal efficiency.

\section{Optimization studies}

Full results of the optimization studies can be found in Ref.~\cite{ourpaper}. We studied the effects of the NC background, the intrinsic beam background, the values of the systematic errors, the influence of the spectral information on the results, the ratio of $\nu:\bar{\nu}$ running, and the possibility of tau detection. The intrinsic beam background is only relevant in case of small $\theta_{13}$ ($\sin^22\theta_{13}<10^{-2}$). The spectral information is vital for improving the CP discovery potential: quasi-elastic events are mostly relevant in the region $\sin^22\theta_{13}\gtrsim 10^{-2}$, while deep inelastic events are vital in the region where $\sin^22\theta_{13}\lesssim 10^{-2}$. For high energy configurations of the beam, we find that equal $\nu$ and $\bar{\nu}$ running times give better results than asymmetric configurations. Finally, tau detection is not only very challenging, but it does not significantly improve the sensitivity to oscillation parameters for any of our setups.

\begin{figure}[htbp]
\begin{tabular}{cc}
\includegraphics[scale=0.35]{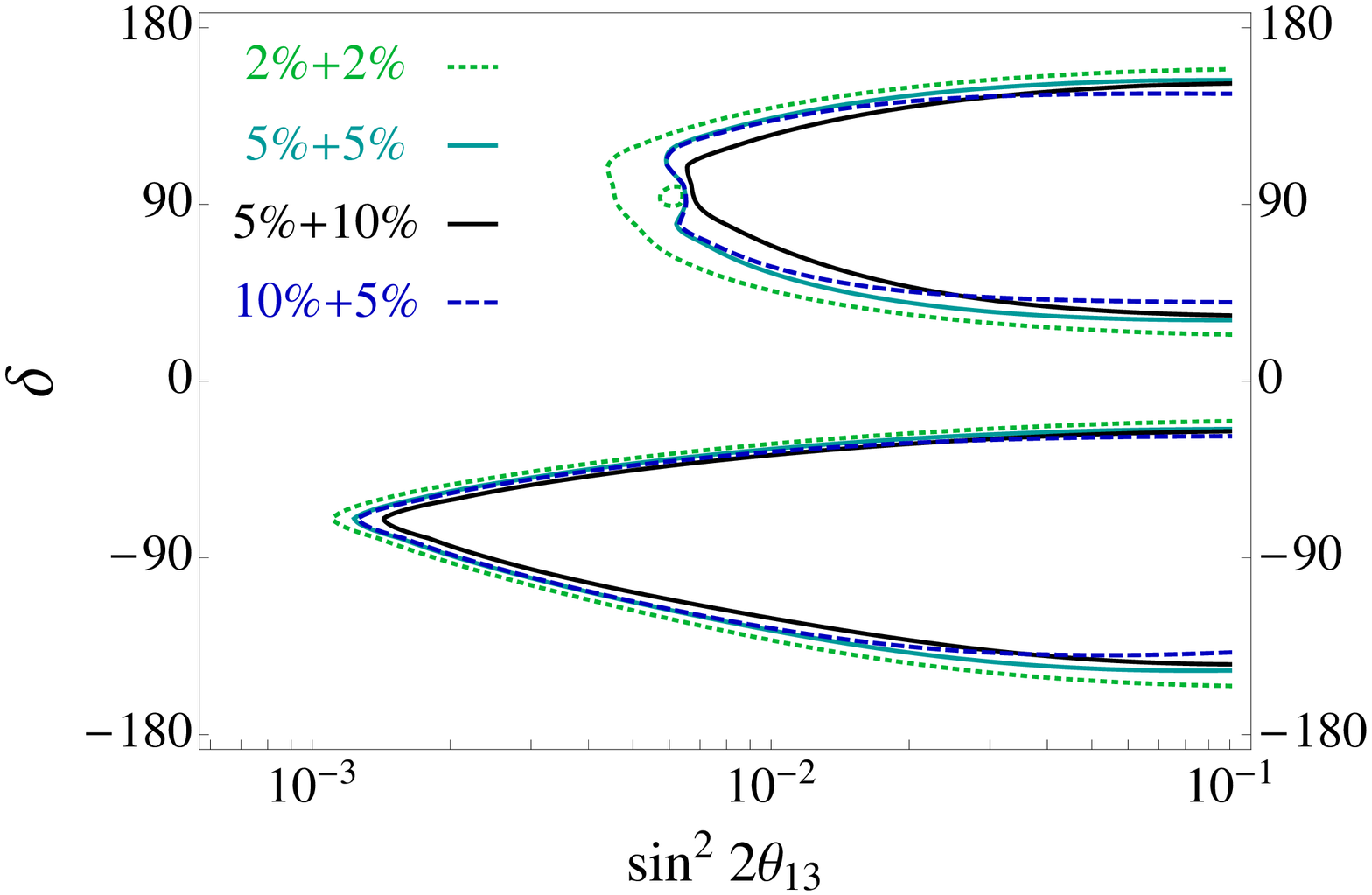} &  
\includegraphics[scale=0.35]{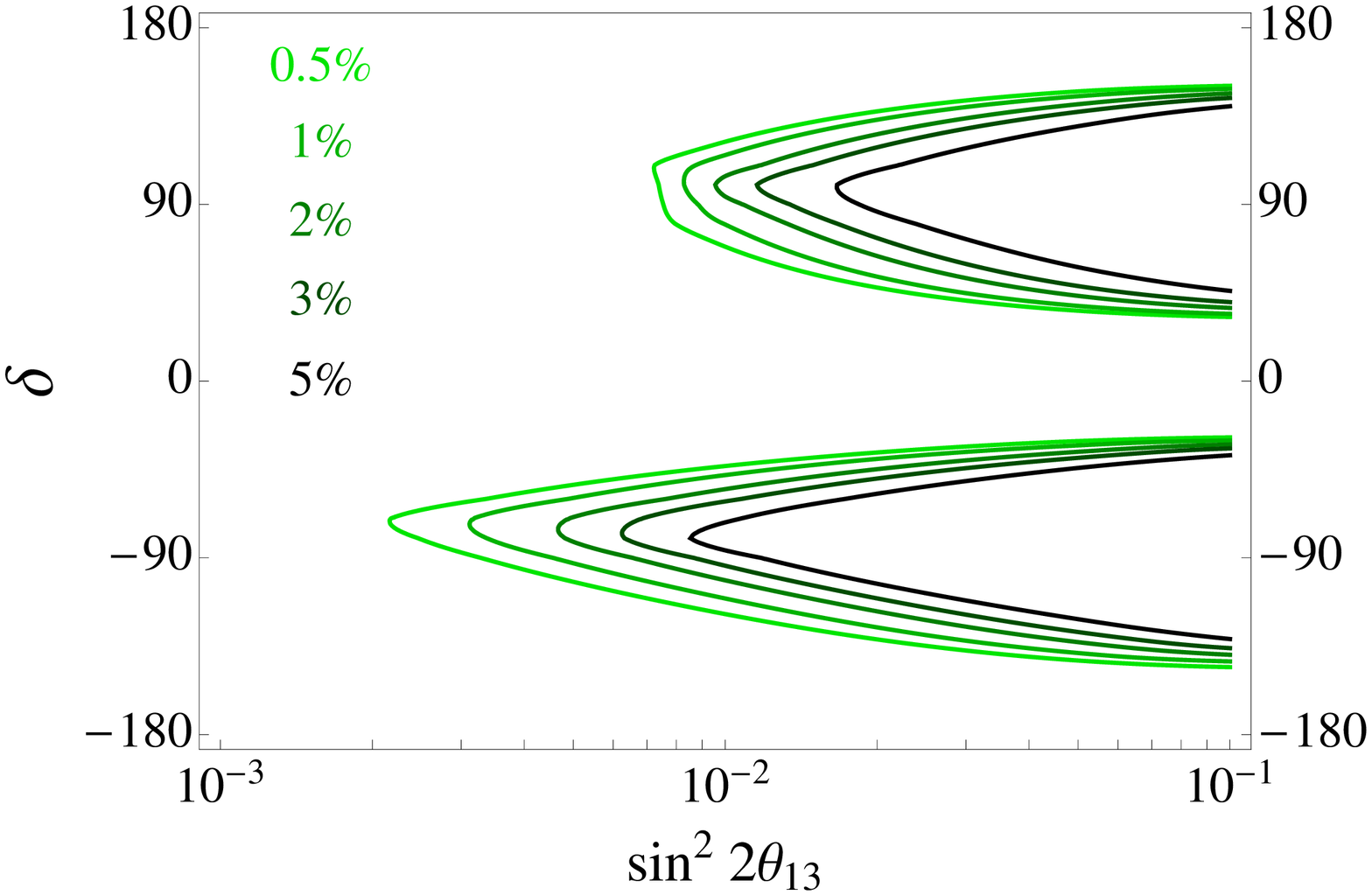}
\end{tabular}
\caption{Effect of systematic uncertainties (left) and the NC backgrounds (right) on the CP discovery potential of a LAr and LS detector, respectively. Results are shown for the detector placed at Pyh\"asalmi. In the region to the right of each line, the CP conservation hypothesis ($\delta=0,\pi$) can be excluded at $3\sigma$ (1 d.o.f.). 
\label{fig:sys}}
\end{figure}

In Fig.~\ref{fig:sys}, we show our results for the impact of systematic errors (left panel) and the NC background (right panel) on the CP discovery potential for the LAr and LS detectors, respectively. Results are shown for the Pyh\"asalmi baseline, but a similar dependence is expected for the rest of locations. In the left panel, the first (second) value in the legend corresponds to the systematic uncertainty over the signal (background), which are taken to be completely uncorrelated. It can be seen that, while the signal systematics is mostly relevant in the region where $\sin^22\theta_{13}\gtrsim 10^{-2}$, background systematics are only relevant below $\sin^22\theta_{13}\sim 10^{-2}$. In the right panel, the value in the legend indicates the percentage of the toatl number of unoscillated NC events that are misidentified as CC events. It can be seen that the results in this case are worsened around a factor of 2 when the NC background is increased from a $0.5\%$ to a $5\%$.

\section{Comparison of baselines and detectors}
\label{sec:results}

In the following, we compare the performance for the different baselines and detectors under consideration. We do not display the results for both the 630 km and 665 km baselines since the results obtained are virtually identical; therefore the results labelled `665' should be understood to also apply to the 630 km baseline. It should be noted that, in the absence of a proper simulation of the WC detector when exposed to a SB with mean neutrino energy around 1.5 GeV, we have used the same migration matrices and rejection efficiencies as for the higher energy setups. These matrices were obtained for a multi-GeV SB and therefore may be pessimistic to describe the performance of the WC for the setups with $L=630$ and $665$ km. This should be taken into account when comparing the results obtained for different baselines. We have also considered a range of values for the NC background in the LS detector. The results for the LS detector in the optimistic scenario (with $0.5\%$ of NC events as background) are very similar to those obtained for the LAr detector and therefore will not be shown here. As it was shown in the previous section, the results for the LS in the pessimistic scenario where a $5\%$ of the NC events are misidentified as CC events are generally around a factor of 2 worse than the ones for the LS with $0.5\%$ NC background, and can be found in Ref.~\cite{ourpaper}.

\textbf{Results for the $\theta_{13}$ discovery potential:} We find that the most relevant factor for this observable is the energy of the beam. Higher neutrino energies imply larger cross sections and, therefore, a larger number of events at the detector. Besides, the $\theta_{13}$ discovery potential of a given facility is not so wildly affected by the presence of degeneracies as the CP or the mass hierarchy discovery potentials. Therefore, in this case we find that the results are usually better for longer baseline setups, for which the spectrum peaks at higher energies. The only exception to this rule is the case of the WC placed at $L=130$ km. In this case, the huge fiducial mass of the detector and its excellent performance in the sub-GeV regime overcome the statistical limitations, giving excellent results for this observable. We find that for the LAr, the WC and the LS with 0.5\% NC background a non-vanishing $\theta_{13}$ can be discovered at the $3\sigma$ CL if $\sin^22\theta_{13}\gtrsim 6\times 10^{-3}$, for any value of $\delta$, with the exception of the setups with LAr or LS placed at $L=130$ km, for which this value is shifted to $\sin^22\theta_{13}\gtrsim 2\times 10^{-2}$. 

\begin{figure}[htbp]
\begin{tabular}{cc}
\includegraphics[scale=0.35]{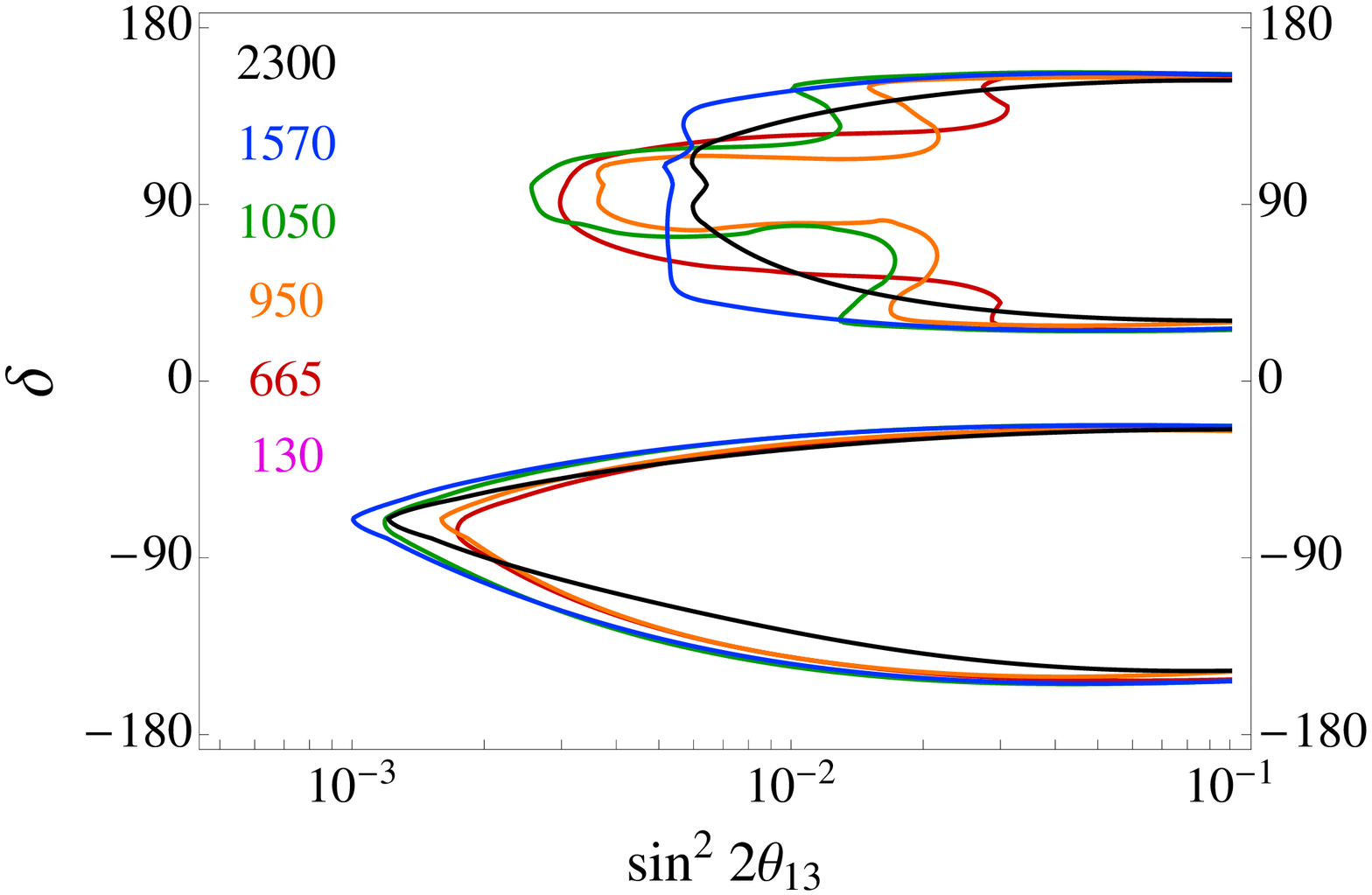} & 
\includegraphics[scale=0.35]{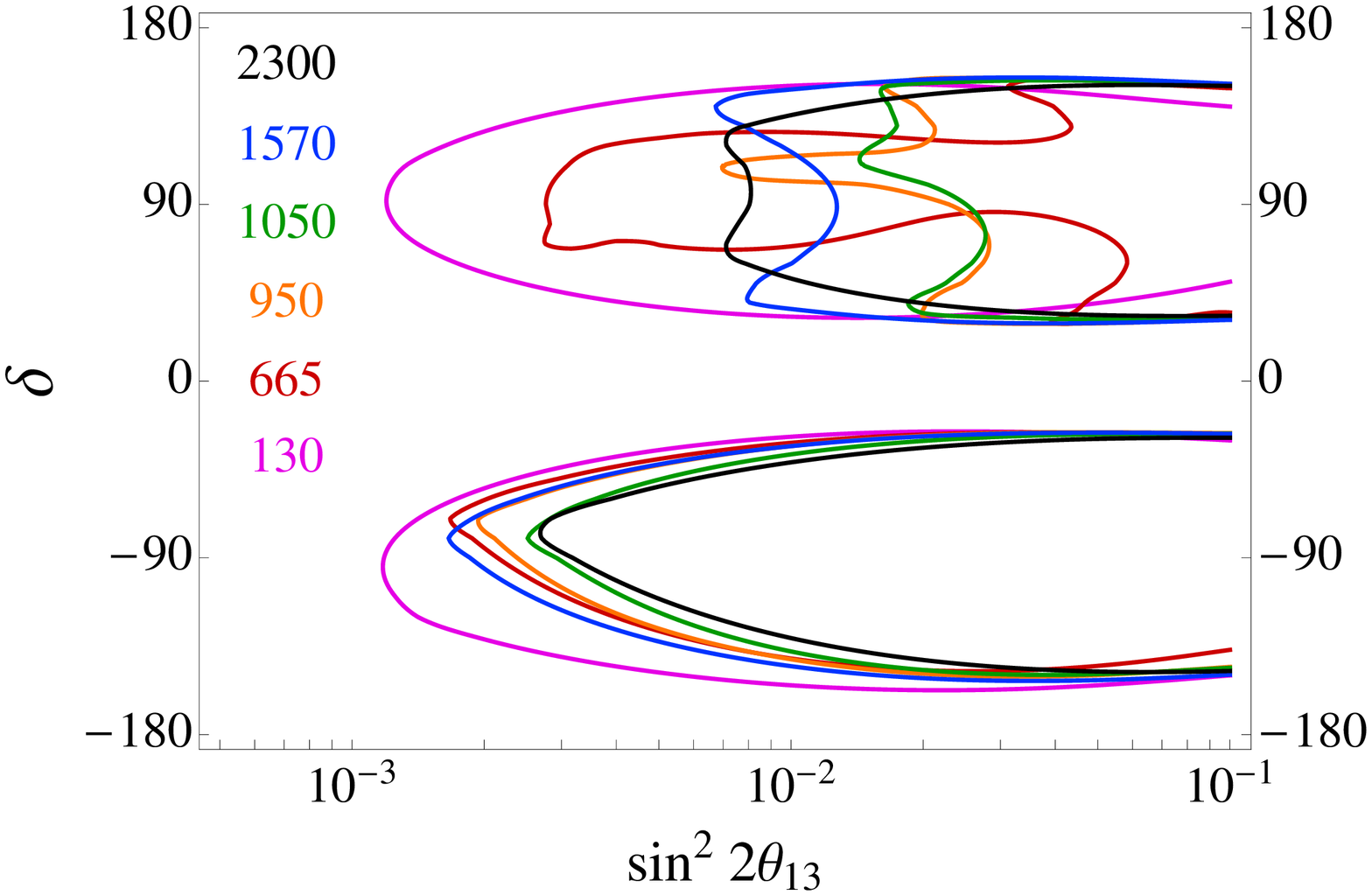}
\end{tabular}
\caption{CP discovery potential for the LAr (left) and the WC (right) detectors, for each of the baselines indicated in the legend. In the region to the right of each curve the CP conservation hypothesis ($\delta=0$ or $\pi$) can be excluded at $3\sigma$ (1 d.o.f.). 
\label{fig:cp}}
\end{figure}

\textbf{Results for the CP discovery potential:} In Fig.~\ref{fig:cp} we show the results for the $3\sigma$ CP discovery potential, for the LAr and the WC detectors.  The baseline dependence is similar to that of the $\theta_{13}$ discovery due to the effect of larger cross sections at higher neutrino energies. However, the CP discovery potential is more affected by degeneracies: this can be clearly observed in the region where $\delta \sim +90^\circ$, where matter effects move the sign degeneracies to CP-conserving values of $\delta$ and therefore the CP discovery potential is severely worsened. Finally, as it can be seen from the figure, the best results for this observable are obtained for the WC detector placed at 130 km from the source. This is again due to its excellent performance in the low energy regime, together with its much larger fiducial mass. In addition, the absence of matter effects reduces the effect of degeneracies and the CP discovery potential is practically unaffected. It should be noticed that both the LAr and the LS detectors give null results for this observable when they are placed at 130 km, as it can be seen from the left panel in the figure. 

\textbf{Results for the mass hierarchy discovery potential:} We find that the length of the baseline is crucial in order to determine the mass hierarchy, regardless of the detector technology. The best results are obtained for the LAr, WC and LS with $0.5\%$ NC background placed at 2300 km: the sensitivity to the mass hierarchy in this case reaches $\sin^{2}2\theta_{13}\sim 8\times10^{-4}$ if $\delta\sim -140^{\circ}$ and $\sin^{2}2\theta_{13}\sim 5 \times 10^{-3}$ if $\delta=+90^{\circ}$. The results for the second baseline in length, $1570$ km, are very close to these and reach $\sin^{2}2\theta_{13}\sim 1.5 \times10^{-3}$ if $\delta \sim -140^{\circ}$ and $\sin^{2}2\theta_{13}\gtrsim 6\times 10^{-3}$ if $\delta=+90^{\circ}$. The results obtained at $665$ and $650$ km are around an order of magnitude worse than these, and the rest of baselines lie in between. Finally, the hierarchy discovery potential is zero for any of the detectors placed at $L=130$ km due to the absence of matter effects.

\section*{Acknowledgements}
This work was funded by the European Community under the European Commission Framework Programme 7 Design Study LAGUNA (Project Number 212343). PC acknowledges financial support from Comunidad Aut\'onoma de Madrid, and from projects HEPHACOS S2009/ESP-1473 (CAM, Spain) and FPA2009-09017 (DGI del MCyT, Spain).

\section*{References}

\bibliographystyle{unsrt}

\end{document}